\documentclass[12pt]{article}
\pdfoutput=1
\usepackage{amssymb,subfigure,graphics,graphicx,bm,cite}
\usepackage[bf,footnotesize,labelsep=period]{caption}
\begin{document}

\title{Wealth distribution across communities \\ of adaptive financial agents\thanks{Published in New Journal of Physics, Volume 17, on August 3rd 2015. P. DeLellis, F. Garofalo, F. Lo Iudice, and E. Napoletano are with the Department of Electrical Engineering and 
Information Technology. Email:\{pietro.delellis\}\{franco.garofalo\}\{francesco.loiudice\}\{elena.napoletano\}@unina.it}}
\author{Pietro DeLellis\thanks{Corresponding author}, Franco Garofalo\\Francesco Lo Iudice, Elena Napoletano}%\corauthref{cor}}
\maketitle
\begin{abstract}
This paper studies the trading volumes and wealth distribution of a novel agent-based model of an artificial financial market. In this model, heterogeneous agents, behaving according to the Von Neumann and Morgenstern utility theory, may mutually interact. A Tobin-like tax (TT) on successful investments and a flat tax are compared to assess the effects on the agents' wealth distribution. We carry out extensive numerical simulations in two alternative scenarios: i) a \textit{reference scenario}, where the agents keep their utility function fixed, 
and ii) a \textit{focal scenario}, where the agents are adaptive and self-organize in communities, emulating their neighbours by updating their own utility function. Specifically, the interactions among the agents are modelled through a directed scale-free network to account for the presence of community leaders, and the herding-like effect is tested against the reference scenario.
We observe that our model is capable of replicating the benefits and drawbacks of the two taxation systems and
that the interactions among the agents strongly affect the wealth distribution across the communities. Remarkably, the communities benefit from the presence of leaders with successful trading strategies, and are more likely to increase their average wealth. Moreover, this emulation mechanism mitigates the decrease in trading volumes, which is a typical drawback of TTs.\end{abstract}

% Uncomment for PACS numbers
%\pacs{00.00, 20.00, 42.10}
%
% Uncomment for keywords
%\vspace{2pc}
%\noindent{\it Keywords}: XXXXXX, YYYYYYYY, ZZZZZZZZZ
%
% Uncomment for Submitted to journal title message
%\submitto{\JPA}
%
% Uncomment if a separate title page is required
%\maketitle
% 
% For two-column output uncomment the next line and choose [10pt] rather than [12pt] in the \documentclass declaration
%\ioptwocol
%

\section{Introduction}

Neoclassical economics plays a fundamental role in the study of price and income distributions in markets \cite{ba:85}. Recent studies are focused on the development of tools and approaches that may complement neoclassical economics, removing some of its main assumptions, such as rationality and homogeneity of the financial agents \cite{make:11,crpi:11}. A special interest emerged for a complex system approach, which involves the use of agent-based and behavioural models \cite{alcr:09,caga:99,papo:10,cifo:03,crpi:11,fafo:09,gi:08,le:01,le:02,make:11,tuli:12,we:10,sajo:12,wedi:06,didi:09,chch:13,pato:13,gu:09,chch:09}. In this way, the simulated behaviour of a high number of decision-makers and institutions (the agents), interacting through prescribed rules, can be investigated to highlight the macro features of the market emerging from the dynamic interactions among the players.

In the last decades, notable artificial financial models have been developed. The model in \cite{le:02} provides a tenable explanation for bubbles and crashes, replicating liquidity crashes that cannot be predicted with equilibrium models. The studies in \cite{luma:99,zaka:02} support the idea that scaling and fluctuations in financial prices arise from mutual agents' interactions rather than from the efficient market hypothesis.
 The dynamics of companies' growth and decline have been reproduced in \cite{ax:01}, reflecting the power-law distribution of company size. In agent-based financial markets, agents can be modelled as zero-intelligence traders, acting randomly with a budget constraint \cite{gosu:93}, or as heterogeneous agents with bounded rationality, endowed with utility functions, who can envision strategies with limited complexity \cite{le:01,am:08,cifo:03}.
%The authors of \cite{cifo:03} considered random agents in a multi-asset financial market, and investigated three alternative trading strategies in two different market conditions. Utility functions are used instead in \cite{lyta:12} to model the behaviour of financial agents in both deterministic and stochastic settings.
Moreover, agents may have the capability of adapting their actions according to their current state, the information gathered from the environment \cite{limi:15}, and the rules governing their behaviour \cite{bale:10}. Different types of adaptive behaviour may be considered, such as learning, search and imitation \cite{blea:92}. Indeed, the capability of an investor of imitating the actions of others is one of the key elements of the widespread phenomenon of herding in financial markets \cite{shbe:14,cigu:09,limo:08,delo:02}.
%The authors of \cite{delo:02} define ``herd behaviour" as the capability of an investor of imitating the actions of other investors. In other words, an agent ``herds when he takes an action he would not have taken had he not known that other investors have taken it" \cite{delo:02}. A lot of studies \cite{shbe:14} show how herding is a widespread phenomenon in financial markets.
% This adaptive behaviour may be modelled through time-varying utility functions. As explained in \cite{cyde:79}, utility functions may change as a result of actual experience.
The concepts of learning and adaptation are also applied to utility functions, that may not be fixed, as observed in \cite{coax:84,cyde:79,delo:02}.

% Complementary studies \cite{pasi:07,lima:00,waya:09,qizh:06,gopl:99} focused on the analysis of market volatility, unveiling the mechanism of price dynamics by analysing  real financial databases. 
%
%Recently, the interaction among the agents has been modelled using tools from complex networks and graph theory, see \cite{bate:14} and references therein. 

Agent-based approaches have also been used to test the effects of policies, regulations and taxation systems on the market dynamics, see for instance \cite{wedi:06,didi:09}. Inspired by the seminal work of Tobin \cite{to:78}, several taxes on financial transactions were proposed to regulate the markets, whose effects have been controversial \cite{hahu:10,lita:13}. {The effect of taxation was also investigated in the kinetic exchange market models, see \cite{chch:13,pato:13,gu:09,chch:09} and references therein. In \cite{chch:13,pato:13}, the authors reviewed the recent advances in the study of income and wealth distribution. In particular, they focused on the kinetic exchange market models and illustrated their microeconomic foundations. 
In \cite{gu:09}, an ad hoc stochastic asset evolution was considered which explained the gamma-function-like income distribution. Later, Ref. \cite{chch:09} provided a microeconomic framework to derive asset equations consistent to those used in the ideal gas market models: at each time step, the agents randomly meet to transact according to the maximization of their Cobb-Douglas utility function. The authors analysed the steady-state income distribution for different values of a redistribution parameter, which mitigates the inequalities in wealth distribution.
In particular, in \cite{chch:09} the steady-state wealth distribution was observed to be gamma-function-like, and studied in presence of a redistributive tax.}

To the best of our knowledge, none of the existing models of artificial markets accounts for herding phenomena and different taxation schemes at the same time. In this work, we build a novel artificial financial market capable of testing the delicate interplay between herding-like interactions, the inequality in wealth distribution, and the balancing of two common alternative taxation schemes. The agents behave according to utility theory, and are grouped in three classes with different risk attitudes and subsequent trading strategies. We study the emerging features of the market in terms of trading volumes and wealth distribution, characterized through the Gini coefficient \cite{gi:12}, in presence of a Tobin-like tax (TT) and a flat tax (FT), respectively. Specifically, we investigate two agents' behavioural scenarios: i) stubborn agents \cite{pobo:07}, who keep their own risk attitude regardless of the effectiveness of the consequent trading strategy (\textit{reference scenario}); ii) adaptive agents, who self-organize in communities to emulate the strategy of their richest neighbours  (\textit{focal scenario}). These latter act as leaders and are modelled as the hubs of a directed scale-free network \cite{baal:99}.

The outline of the paper is as follows. In section \ref{sec:model}, we describe the proposed model of a financial market in terms of agents behaviour, alternative taxation schemes, and interaction rules. In section \ref{sec:results}, the focal scenario is tested against the reference one. The interplay between the different taxation systems and agents' behavioural adaptation is analysed, with a special emphasis on the effects on wealth distribution and trading volumes. Finally, conclusions are drawn in section \ref{sec:concl}.
\section{\label{sec:model}The model}

We introduce an agent-based model of a financial market populated by a set of $n$ agents, who can choose among alternative financial assets. The state of each agent $j$ is defined as his current wealth $x_j$ and risk attitude $\alpha_j$. The agents behave according to the Von Neumann and Morgenstern utility theory \cite{vomo:07}, and alternative taxation schemes and interaction rules are considered.
\subsection{\bf {Market structure}}
At time $0$, the state of agent $j$ is given by his initial wealth and risk attitude, denoted $x_{j0}$ and $\alpha_{j0}$, respectively, $j=1,\ldots,n$.
At each time step $k=1,2,...$, a simulated trading session is performed. Each agent, in a sequential random order, evaluates the convenience of investing a given fraction $\delta$ of his current wealth $x_j(k)$ in one of the financial assets from the set $\mathcal{I}=\{1,\ldots,m\}$. The assets in $\mathcal{I}$ are characterized by a limited availability $A_i, i=1,...,m$, where $A_m = +\infty$ is associated to a virtual asset, corresponding to no-investment. In view of this, each agent is allowed to invest in one of the available assets, that is, in any element of $\mathcal{I}$ such that $A_i \ge \delta x_j(k)$. Agents' access to trading is randomly permuted at each time step $k$, so that, on average, no agent is favoured. After each trading, the availability of the selected asset is updated before the next agent is allowed to trade.\subsection{\bf {Agent behaviours}} 

\subsubsection{Agent trading strategy} A power-law utility function characterizes the risk attitude of each agent. 
%We select the power-law function
%\begin{equation}
%U_j(r_i(k))=r_i(k)^{{\alpha}_j(k)},
%\label{eq:1}
%\end{equation}
%where
%$r_i(k)$ is the expected return of the \textit{i}-th asset at time $k$ and 
%${{\alpha}_j}(k)$ is a (possibly time-varying) coefficient which determines the risk attitude of the \textit{j}-th agent. 
At each trading session $k$, agent $j$ decides to invest a
%the opportunity of investing the 
fraction $\delta$ of its current wealth $x_j(k)$ in the most profitable asset $i \in \mathcal{I}$, selected by comparing the expected utilities
\begin{equation}
E[U_j(x_j(k),i)]=0.5\left[(a_i \delta x_j(k))^{\alpha_j(k)}+(b_i \delta x_j(k))^{\alpha_j(k)}\right], \ i=1,...,m,
\label{eq:utility}
\end{equation}
%\begin{align}
%& U_j(x_j(k),i)= \nonumber \\
%& \left\{\begin{array}{ll}
%0.5\left[(a_ix_j(k))^{\alpha_j(k)}+(b_i\x_j(k))^{\alpha_j(k)}\right] \ i=1,\ldots,m-1, 
%\end{array}\right.\label{eq:utility}
%\end{align}
where ${{\alpha}_j}(k)$ is a (possibly time-varying) coefficient which determines the risk attitude of the \textit{j}-th agent, $a_i$ and $b_i$ are the win and loss rates associated to the \textit{i}-th asset, $i=1,\ldots,m$ \footnote{Notice that the win and loss rates associated to the virtual asset are $a_m = b_m = 1$.}. 
Based on their current risk attitude, we group the agents in three classes. In the first one, there are the agents characterized by a low risk attitude, denoted in what follows as {\it prudent} agents. The agents that are more prone to take risks are denoted {\it audacious} and grouped in the third class. Finally, the intermediate class groups the {\it ordinary} agents.
We emphasize here that an agent may decide not to invest (formally, to invest in the $m$-th asset), if $E[U_j(x_j(k),m)] \ge E[U_j(x_j(k),i)]$ for all $i=1,\ldots,m-1$. 
% At each iteration $k$, the \textit{i}-th asset availability $A_i$ is updated by subtracting from the availability itself the wealth fraction invested by the agent $x_j$: 
%\begin{equation}
%A_i(j+1,k)=A_i(j,k)-x_j(k);    j=1,..,N; i=1,...,n-1.
%\end{equation}
The outcome of the trade is the realization $\beta$ of a uniform Bernoulli random variable $B$. Therefore, the wealth $x_j^-(k)$ of the agent $j$ at time $k$ before the taxation is given by: 
\begin{equation}
x^-_j(k)=x_j(k-1)+\beta \delta x_j(k)(a_i-1)-(1-\beta) \delta x_j(k)(1-b_i).
\label{eq:2}
\end{equation}
When the trading session is over, a tax is applied and the wealth of agent $j$ at time $k$ is updated as
\begin{equation}
x_j(k)=\tau(x^-_j(k)),\label{eq:tax}
\end{equation}
where $\tau$ is the function describing the selected taxation scheme, see section \ref{tax:schemes}.
\subsubsection{Interaction among the agents} We consider two alternative scenarios. In the {\it reference scenario}, the market is composed of stubborn agents, who do not modify their utility function even if they observe that their investing strategy is not successful. Accordingly, their risk attitude is considered as a parameter rather than an evolving state, that is, $\alpha_j(k)=\alpha_{j0}$, for all $k\in\mathbb{N}$, $j=1,\ldots,n$. 
In the {\it focal scenario}, instead, the agents are adaptive, as they are prone to directly interact with each other and update their trading strategy. In particular, we model the strategy modification as a variation of the coefficient $\alpha_j(k)$ in (\ref{eq:utility}). We emphasize that $\alpha_j(k)$ is referred to as ``risk attitude" for simplicity, but it may also embed other relevant factors determining the expected utility function of agent $j$, such as the perceived information level of the other investors \cite{delo:02}.
The reciprocal influence among the agents diffuses through a connection topology described by a directed graph $\mathcal G = \{{\mathcal V, \mathcal E}\}$, where $\mathcal V$ is the set of nodes, corresponding to the agents, and $\mathcal E$ is the set of directed edges connecting the nodes. The existence of an edge $(i,j)$ implies that the risk attitude of node $j$ is influenced by that of node $i$. The herding-like dynamics of the coefficients $\alpha_j(k)$ in (\ref{eq:utility}) is described by
\begin{equation}
\alpha_j(k)=(1-w)\alpha_j(0)+\frac{w}{|\mathcal{N}_j|}\sum_{h\in\mathcal{N}_j}\alpha_h(k-1),
\label{eq:5}
\end{equation}
where $w$ is the interaction weight, $\alpha_j(0)=\alpha_{j0}$, and $\mathcal{N}_j$ is the set of the neighbours of the \textit{j}-th agent, defined as $\mathcal N_{j}=\left\{ i\in\mathcal V : (i,j)\in\mathcal E\right\}$. 
%In what follows, we consider {\it emulating the rich} dynamics, where the richest agents are stubborn, but they influence the other agents, so playing the role of \textit{leaders} \cite{elfa:13}.
We remark that the bigger the coefficient $w$ is, the more the agents are prone to modify their utility function: $w=0$ models the case of stubborn agents, while $w=1$ the case in which the agents completely disregard their innate risk attitudes and emulate the neighbours' behaviour.
\subsection{{\bf Leaders and communities}}

%The different risk attitudes of the agents are assumed to create {\it communities} inside the market. Accordingly, 
The interaction topology is modelled as a disjoint directed scale-free network, and the graph $\mathcal{G}$ is decomposed in up to three disconnected components, the {\it communities}, each of which is guided by \textit{leaders} belonging to the same risk attitude class. Namely, inside each community, we consider {\it emulating the rich} dynamics, where the richest agents are stubborn, but influence the other agents, so playing the role of leaders \cite{elfa:13}. {We choose to consider separated communities so that each follower cannot be influenced by leaders with significantly different risk attitudes. Accordingly, each follower elects to emulate the strategy he considers most profitable.} The size of the communities is proportional to the total wealth of their leaders and, inside each community, the richest agents are more likely to activate links.

%The complete structure of the graph $\mathcal{G}$ is established at $k_t$, when the interaction dynamics are triggered. At this time instant, the richest agents are more likely to activate outgoing links. 
The interaction is triggered at a given time instant $k_t$. Henceforth, the dynamics of $\alpha_j(k)$, $j=1,...,n$, described in  (\ref{eq:5}), are strongly influenced by the structure of the graph $\mathcal{G}$ describing the diffusion flow. In turn, the structure of $\mathcal{G}$ is established at time $k_t$, based on the current wealth $x_j(k_t)$, for $j=1,\ldots,n$. 
\subsection{{\label{tax:schemes}\bf Taxation schemes}}

\begin{figure}
\centering
\subfigure[]{\label{Fig1a}\includegraphics[width=6.7cm]{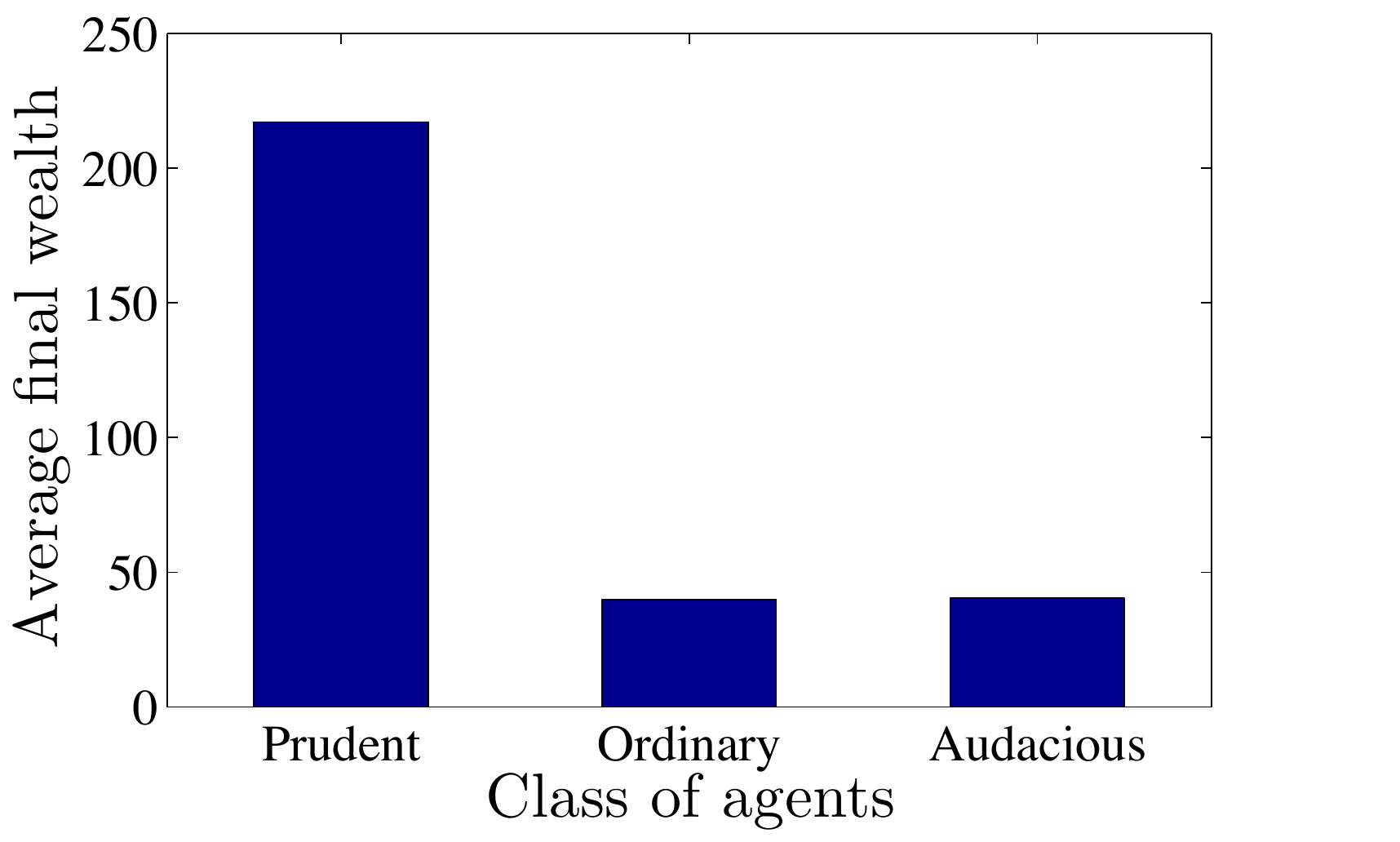}}
\subfigure[]{\label{Fig1b}\includegraphics[width=6.7cm]{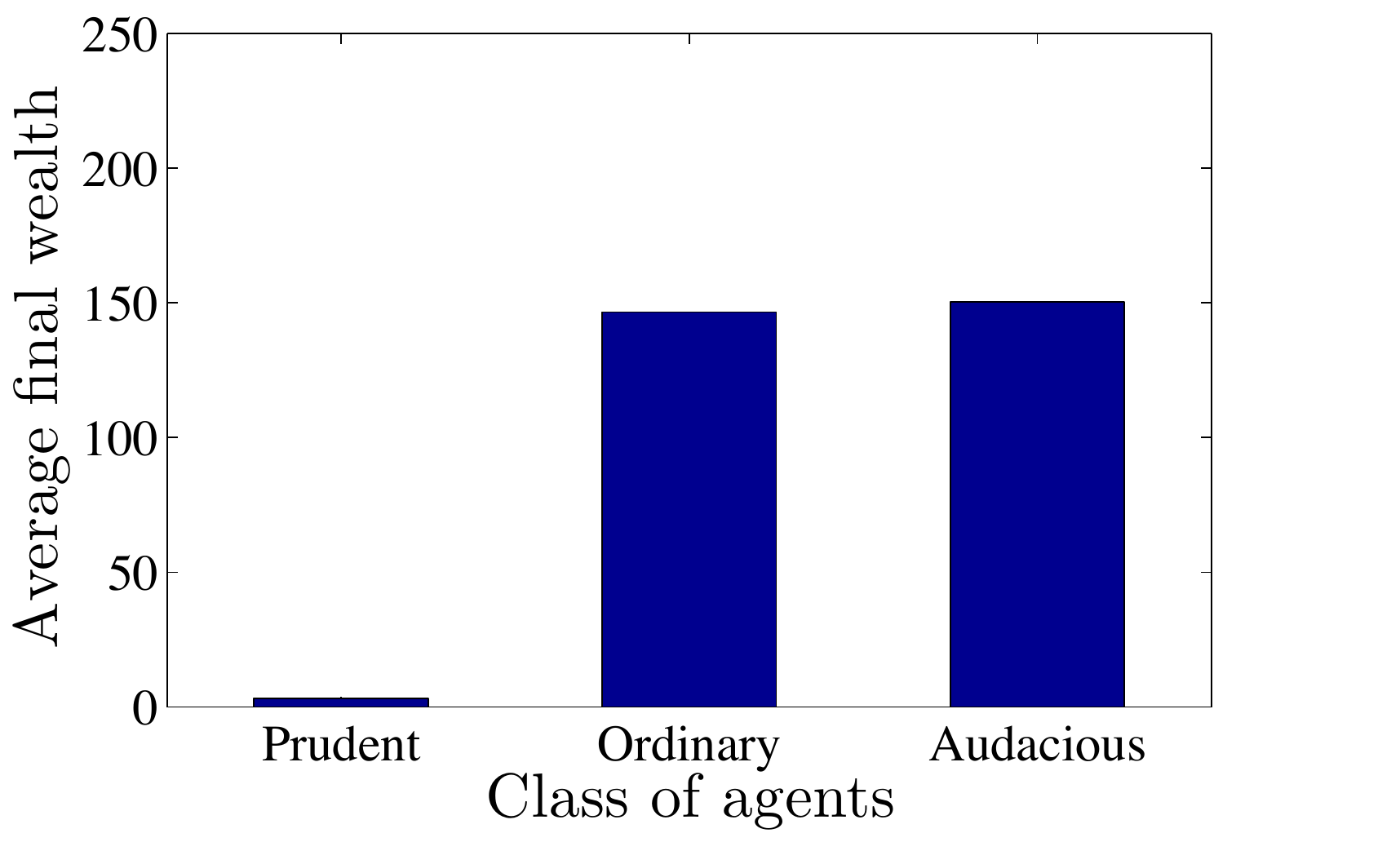}}
\caption{Reference scenario. Final wealth distribution when TT \subref{Fig1a} and FT \subref{Fig1b} schemes are introduced, respectively.}\label{Fig1}
\end{figure}
We consider two alternative taxation systems, which affect the current wealth of the agents $x^-_j(k)$ in different ways: a) taxation on financial transactions, and b) taxation on wealth. 
%Taxes on financial transactions have been adopted in several countries in the last century: a well-known example is the so-called Tobin Tax \cite{to:78}, named after the Nobel prize James Tobin, whose original scope was to put a penalty on short-term financial round-trip excursions into another currency. 
Type a) tax is a Tobin-like tax, which reduces the current wealth of the winning agents by a profit fraction $u(k)$ given by
\begin{equation}\label{eq:h}
u(k)= \left\{\begin{array}{ll}
\frac{p(k)}{\sum_{j=1}^ns_j(k)}, \ \ \,p(k)>0, 
\\ 0,\qquad\qquad p(k)\leq 0,
\end{array}\right.
\end{equation}
where $s_j(k)=x^-_j(k)-x_j(k-1)$, and $p(k)=\sum_{j=1}^n(x^-_j(k)-x_{j0})$. Accordingly, (\ref{eq:tax}) becomes
\begin{equation}
x_j(k)=x^-_j(k)-H(s_j(k))s_j(k)u(k),
\end{equation}
where $H$ is the Heaviside step function.

The alternative taxation scheme b) is a FT, in which the amount of the tax is proportional to the total wealth of the individual. %Specifically, it is a non-progressive wealth tax (WT), proportional to the current wealth $x^-_j(k)$ of each agent $j$, with $j=1,\ldots,n$. 
Accordingly, (\ref{eq:tax}) becomes
\begin{equation}
x_j(k)=v(k)x^-_j(k),
\label{eq:3}
\end{equation}
where  $v(k)=\frac{\sum_{j=1}^nx_{j0}}{\sum_{j=1}^nx^-_j(k)}$. 

Notice that, to allow for a proper comparison between the two taxation schemes, the coefficients $u(k)$ and $v(k)$ in (\ref{eq:h}) and (\ref{eq:3}), respectively, are selected so as to keep the average wealth constant over time, that is, $\frac{1}{n}\sum_{j=1}^nx_j(k)=\bar{x}$.

\begin{figure}
\centering
\subfigure[]{\label{Fig2a}\includegraphics[width=6.7cm]{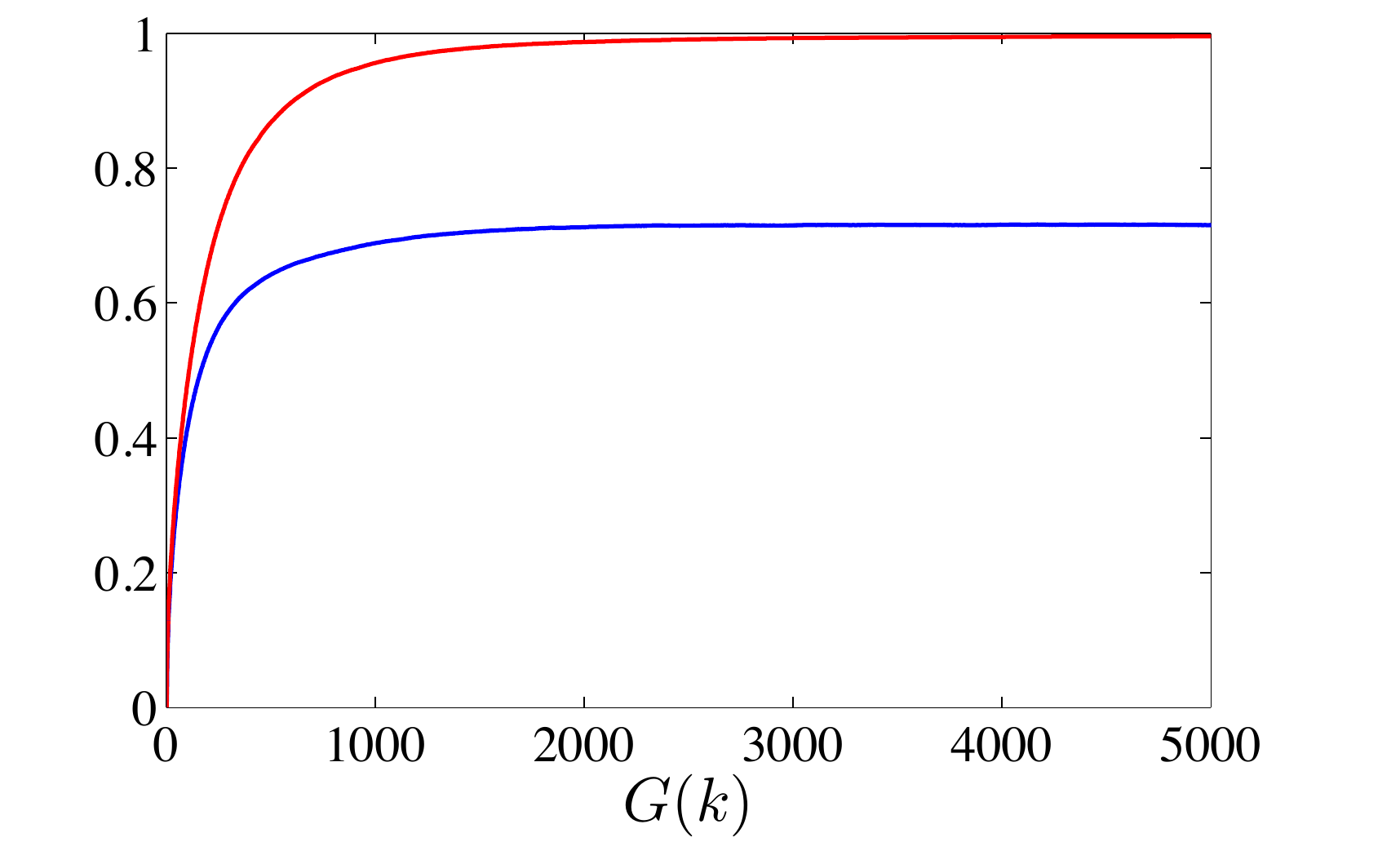}}
\subfigure[]{\label{Fig2b}\includegraphics[width=6.7cm]{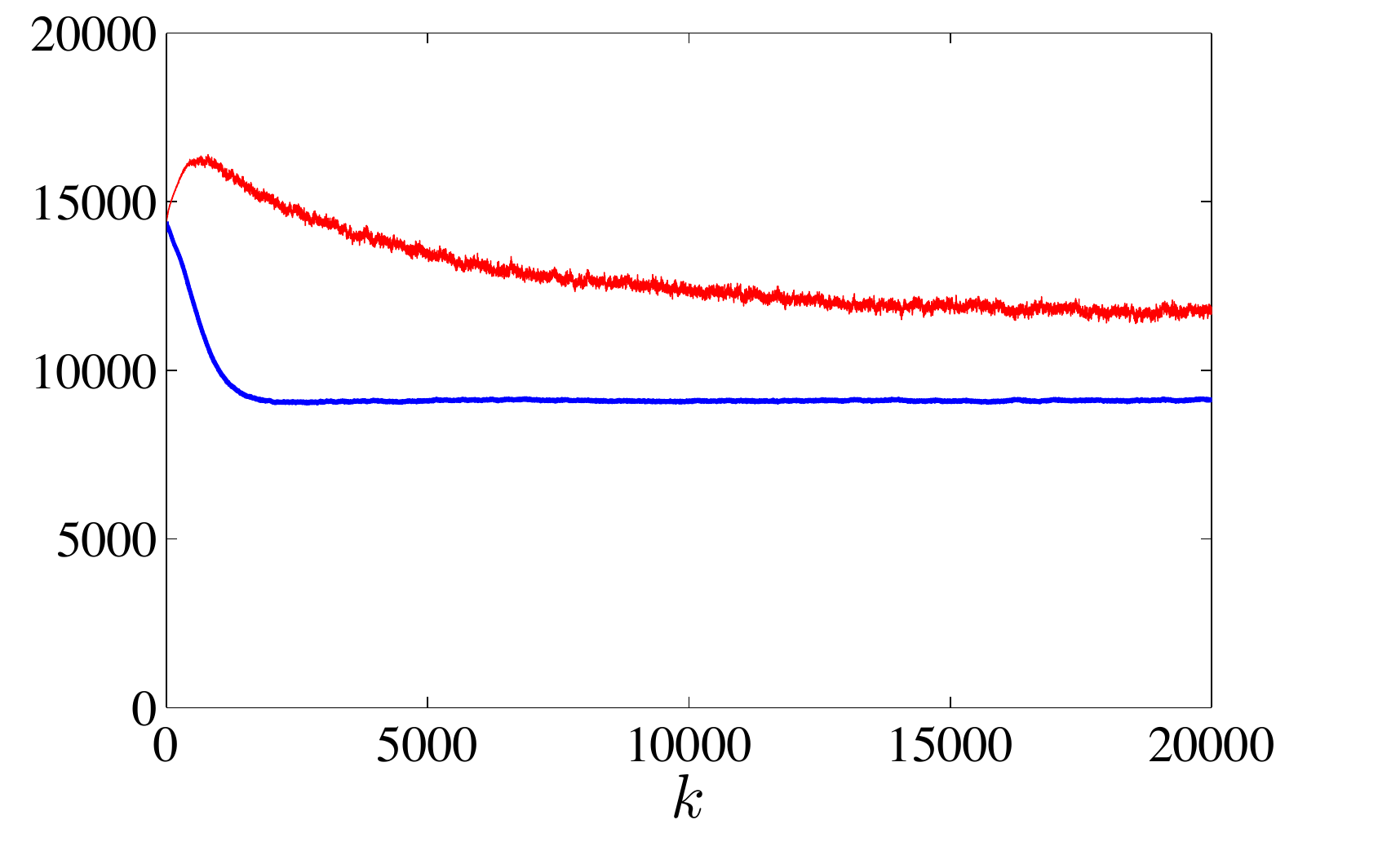}}
\caption{Reference scenario. Gini coefficient \subref{Fig2a} and trading volumes \subref{Fig2b} when TT (blue line) and FT (red line) schemes are introduced, respectively.}\label{Fig2}
\end{figure}
%\begin{figure}
%\centering
%\includegraphics[width=7cm]{fig2}
%\caption{Reference scenario. Gini coefficient when TT (blue line) and FT (red line) schemes are introduced, respectively.}\label{Fig2}
%\end{figure}
%%
%\begin{figure}
%\centering
%\includegraphics[width=7cm]{fig3_mod}
%\caption{Reference scenario. Trading volumes when TT (blue line) and FT (red line) schemes are introduced, respectively.}\label{Fig3}
%\end{figure}

\begin{figure}
\centering
\subfigure[]{\label{Fig9a}\includegraphics[width=6.7cm]{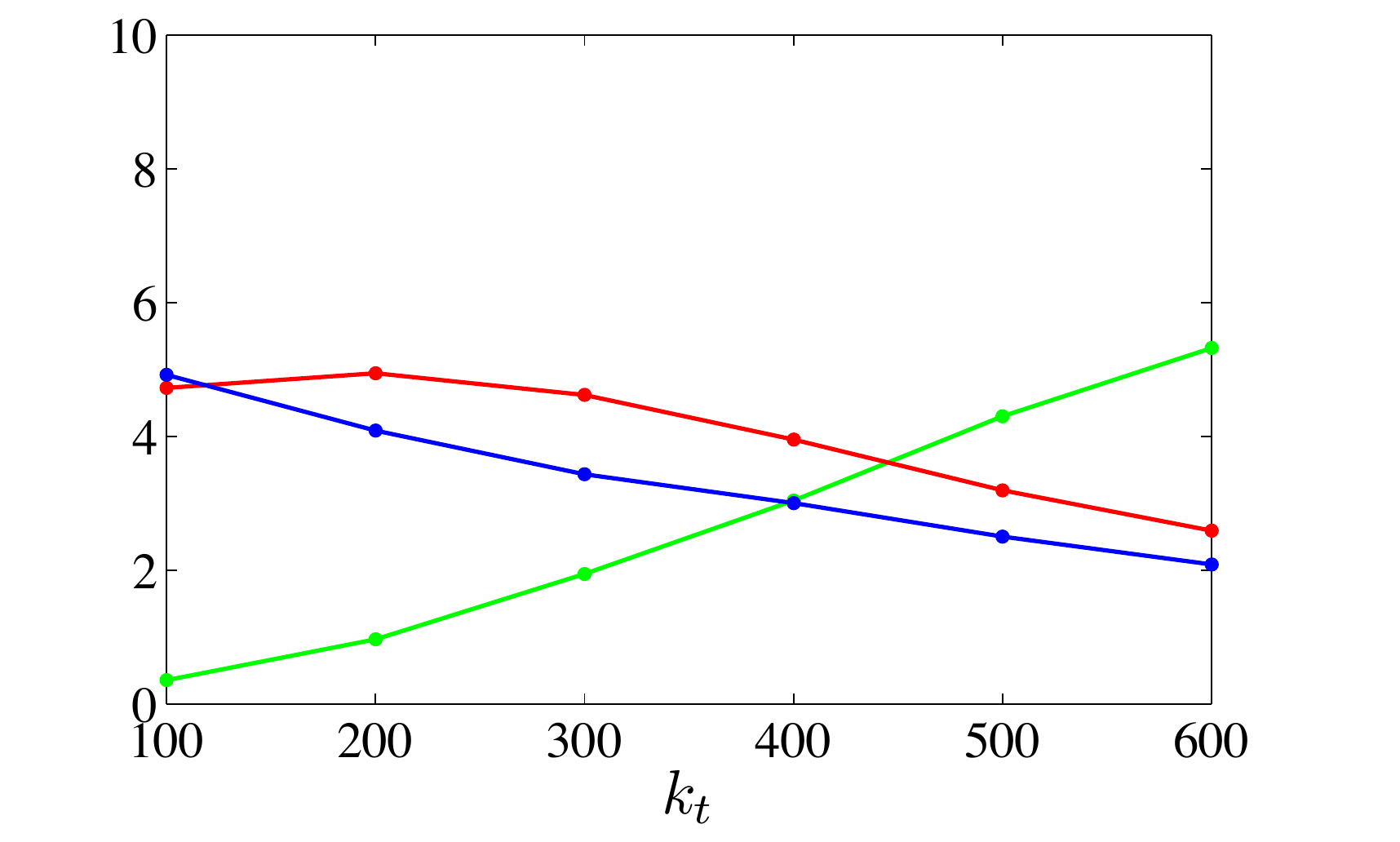}}
\subfigure[]{\label{Fig9b}\includegraphics[width=6.7cm]{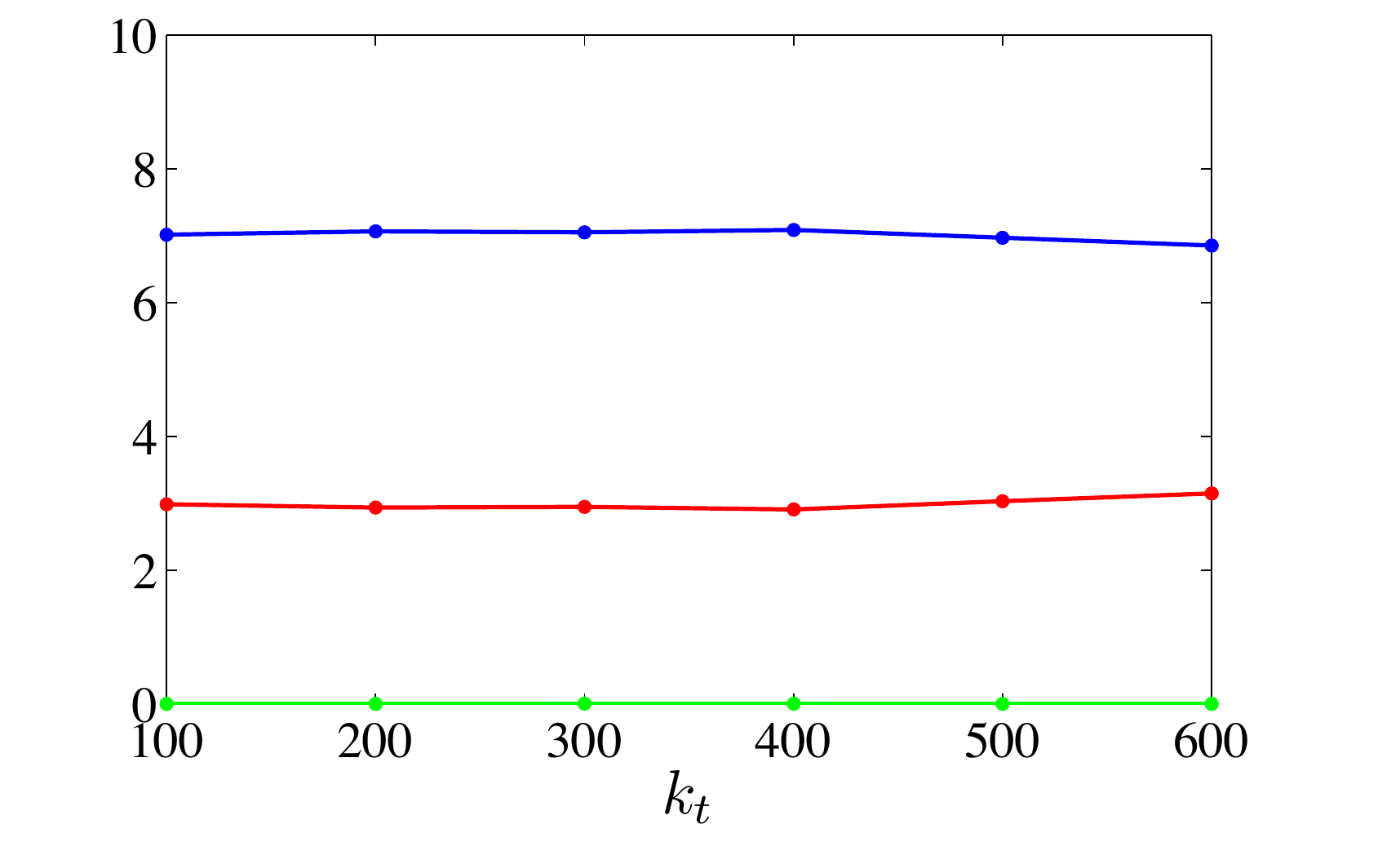}}
\caption{Focal scenario. Average number of leaders of belonging to the communities 1 (red line), 2 (blue line) and 3 (green line) for different interaction triggering time $k_t$ when TT \subref{Fig9a} and FT \subref{Fig9b} schemes are introduced, respectively.}\label{Fig9}
\end{figure}
\begin{figure}
\centering
\subfigure[]{\label{Fig4a}\includegraphics[width=6.7cm]{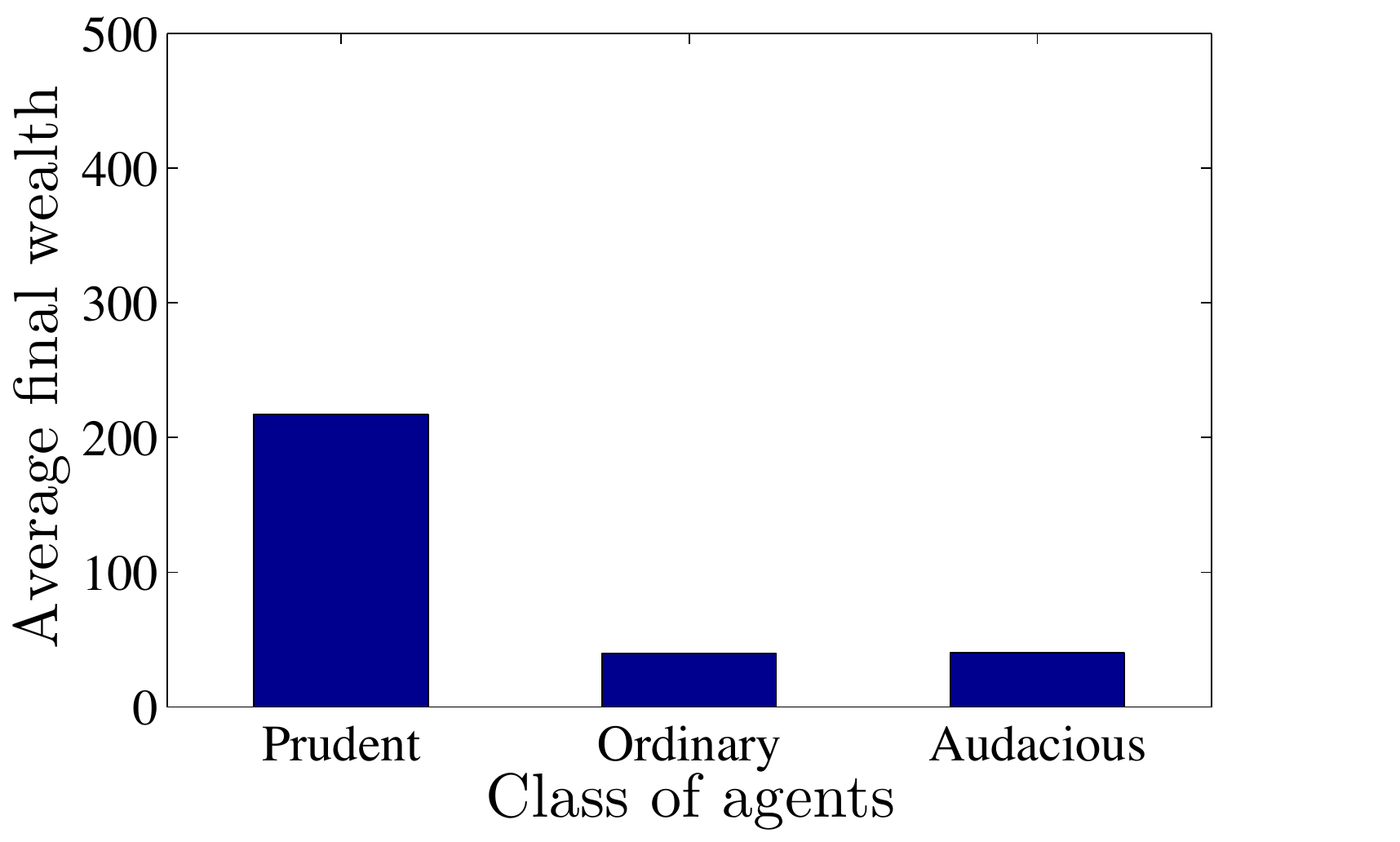}}
\subfigure[]{\label{Fig4b}\includegraphics[width=6.7cm]{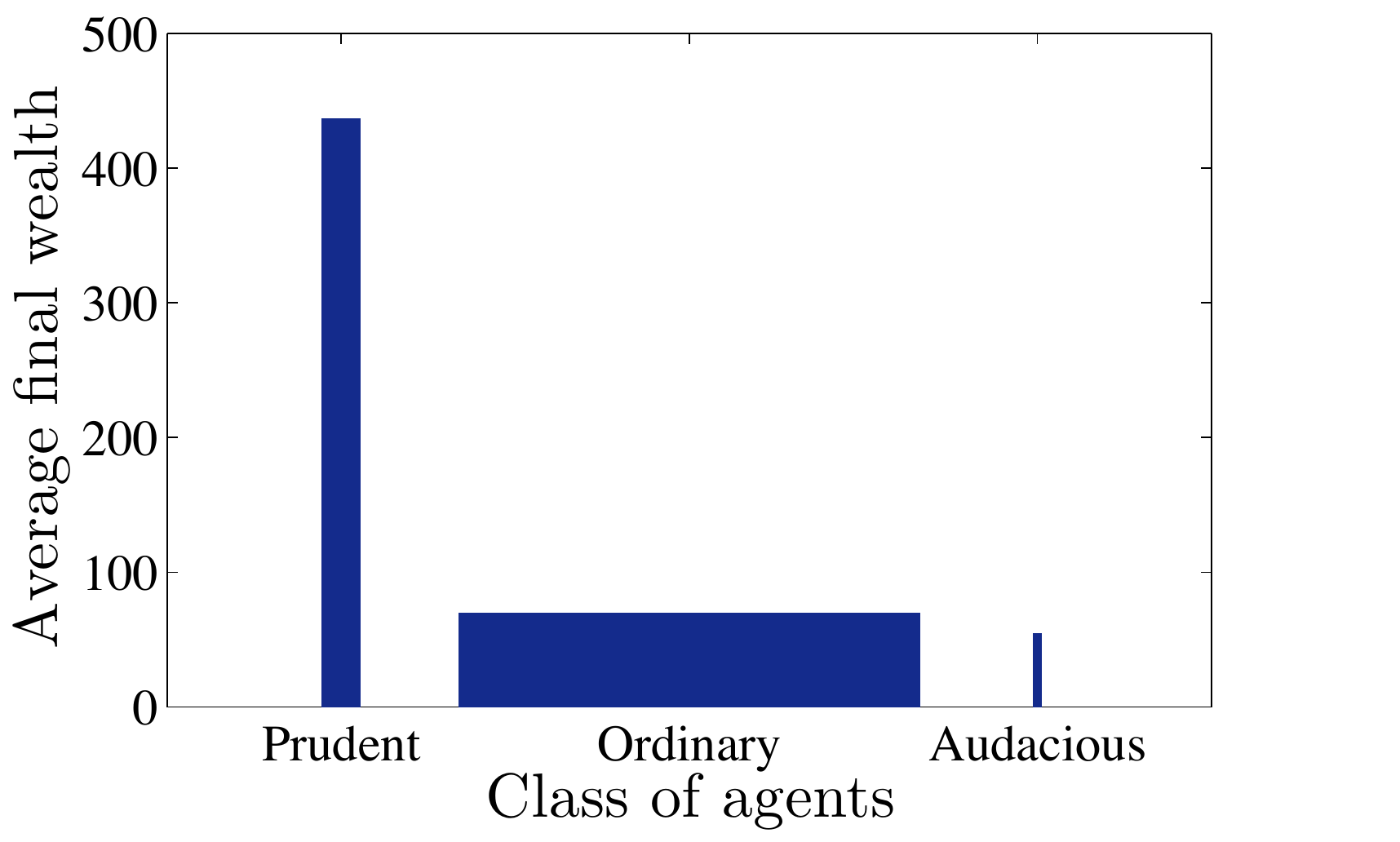}}
\caption{Final wealth distribution when TT scheme is introduced: reference \subref{Fig4a} and focal \subref{Fig4b} scenarios, respectively. The width of the bars is proportional to the numerosity of the classes.}\label{Fig4}
\end{figure}

\section{\label{sec:results}Results}
The proposed model of artificial financial market can take into account different scenarios, in terms of both taxation schemes and interaction rules. In this section, we analyse the emerging features in the focal scenario, aiming at identifying the possible interplay between taxation and interaction in determining the trading volumes and the wealth distribution among the agents. To do so, we compare the results of extensive simulations of the focal scenario against those of the reference one. The effect of the alternative taxation schemes are firstly pointed out in the reference scenario. Then, we focus on adaptive agents and study the effects induced by the emulating dynamics on the wealth distribution for both taxation schemes. To achieve statistical relevance, we run {1000} simulations for each scenario and consider a number of time steps sufficient to reach steady-state wealth distribution.

We consider $n=1000$ agents that share the same initial capital $x_{j0}=100\$$, $j=1,\ldots,n$,  and, at each trading session $k$, can decide to invest a fraction $\delta = 0.2$ of their current wealth. The cardinality of the set of assets  $\mathcal{I}$ is $m=4$, that is, the agents can trade in three categories of actual assets, while the fourth one corresponds to no-investment and, therefore, has an unlimited availability. On the other hand, at every time instant, each of the three actual assets has an availability equal to $1/15$th of the total wealth of the system. The win and loss rates associated to the actual assets are selected so that the prudent agents $(0.5 \leq \alpha_j(k) < 0.67)$ only consider investing in the first and less risky asset, the ordinary agents $(0.67 \leq \alpha_j(k) < 0.83)$ also consider the second, while the audacious agents $(0.83 \leq \alpha_j(k) \leq 1)$ also find convenient investing in the third and most risky one. Namely, the won rates are $a_1=1.53$, $a_2=1.60$, and $a_3=1.67$, while the loss rates are $b_1=0.6$, $b_2=0.5$, and $b_3=0.4$. Notice that, with this parameter selection, $E[U_j(x_j(k),1)]>E[U_j(x_j(k),2)]>E[U_j(x_j(k),3)]$, $j=1,...,n$. 
The initial risk attitudes $\alpha_{j0}$, $j=1,...,n$, are uniformly distributed in the interval $[0.5, 1]$.

\begin{figure}
\centering
\includegraphics[width=\columnwidth]{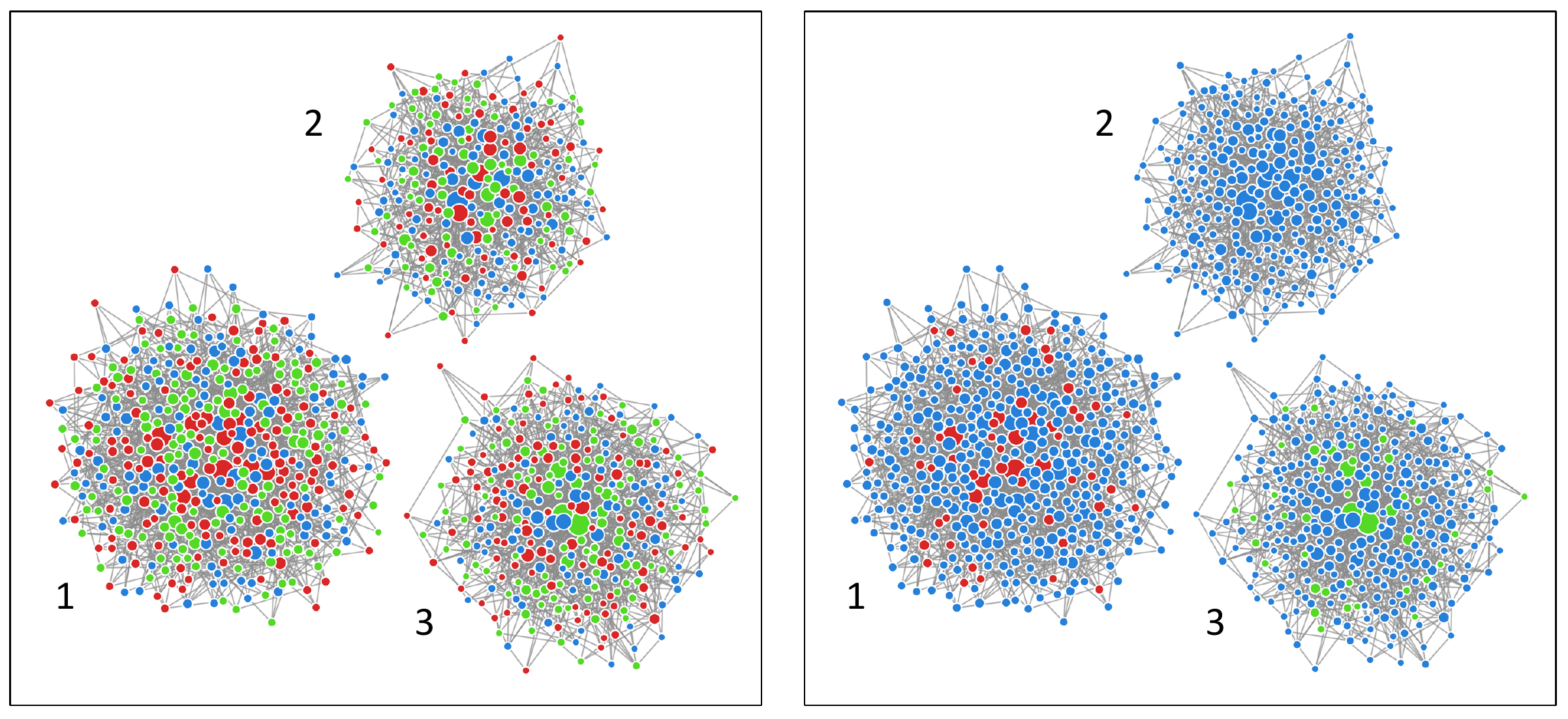}
\caption{Focal scenario, TT scheme. Screenshots of the three communities before (left panel) and after (right panel) the herding-like effect in a sample simulation. The red, blue, and green nodes correspond to prudent, ordinary, and audacious agents, respectively. Notice that the averaging of the attitudes increases the overall density of ordinary agents. However, the leaders' influence skews the distribution across the communities, with prudent and audacious agents still populating the first and third community, respectively.}\label{grafi_v2}
\end{figure}
\subsection{{\bf Reference scenario}} In our numerical analysis, we monitor the effect of the alternative taxation schemes on both the wealth distribution and the trading volumes in the artificial market. Specifically, we observe that the TT scheme hinders the audacious agents, favouring the prudent ones. This is clearly depicted in figure \ref{Fig1}\subref{Fig1a}, which shows the sum of the average final wealth in the three classes of agents, respectively. The opposite is observed when a flat tax is adopted, in which the wealth distribution is biased towards the ordinary and audacious agents, see figure \ref{Fig1}\subref{Fig1b}. In other words, the TT scheme does not reward the risk, penalizing the audacious agents, in opposition to the FT scheme, which encourages the agents to trade.
%

%
%\begin{figure}
%\subfigure[]{\label{Fig3a}\includegraphics[width=7cm]{fig3a}}
%\subfigure[]{\label{Fig3b}\includegraphics[width=7cm]{fig3b}}
%\caption{Trading volumes when TT \subref{Fig3a} and WT \subref{Fig3b} schemes are introduced, respectively.}\label{Fig3}
%\end{figure}
%
A second striking difference among the two taxation schemes is the overall wealth dispersion. To point out this emerging feature, we consider the Gini coefficient $G(k)$, proposed by Corrado Gini in \cite{gi:12} as a measure of inequality of income or wealth, which can be defined as
\begin{equation}
G(k)=1-\frac{2}{n-1}\left( n-\frac{\sum_{j=1}^njx_j(k)}{\sum_{j=1}^nx_j(k)}\right),
\label{eq:Gini}
\end{equation}
where the wealths $x_j(k)$, $j=1,\ldots,n$, are indexed in non-decreasing order $(x_j(k)\leq x_{j+1}(k))$.
 The Gini coefficient varies between 0, which reflects complete equality, and 1, which indicates complete inequality (one person holds the all wealth, all others have none). As depicted in figure \ref{Fig2}\subref{Fig2a}, while the TT scheme induces a wealth redistribution among the population, the FT scheme increases inequalities. On the other hand, the TT scheme leads to lower trading volumes at the steady-state, see figure \ref{Fig2}\subref{Fig2b}. The latter effect is in line with the criticisms commonly made to financial transaction taxes, which are blamed for possible market depression \cite{bali:06,hahu:10,mama:08}.
\begin{figure}
\centering
\subfigure[]{\label{Fig5a}\includegraphics[width=6.7cm]{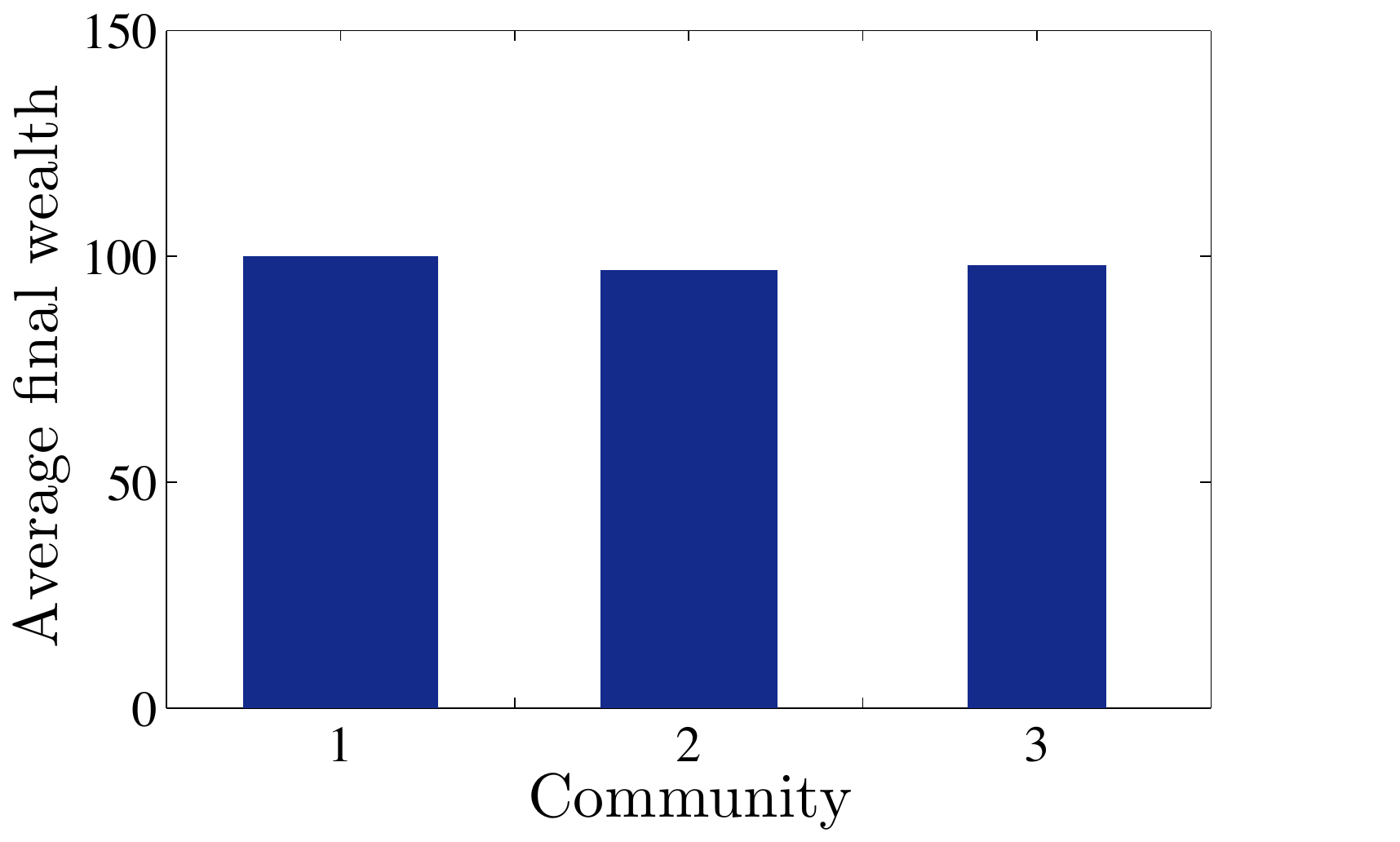}}
\subfigure[]{\label{Fig5b}\includegraphics[width=6.7cm]{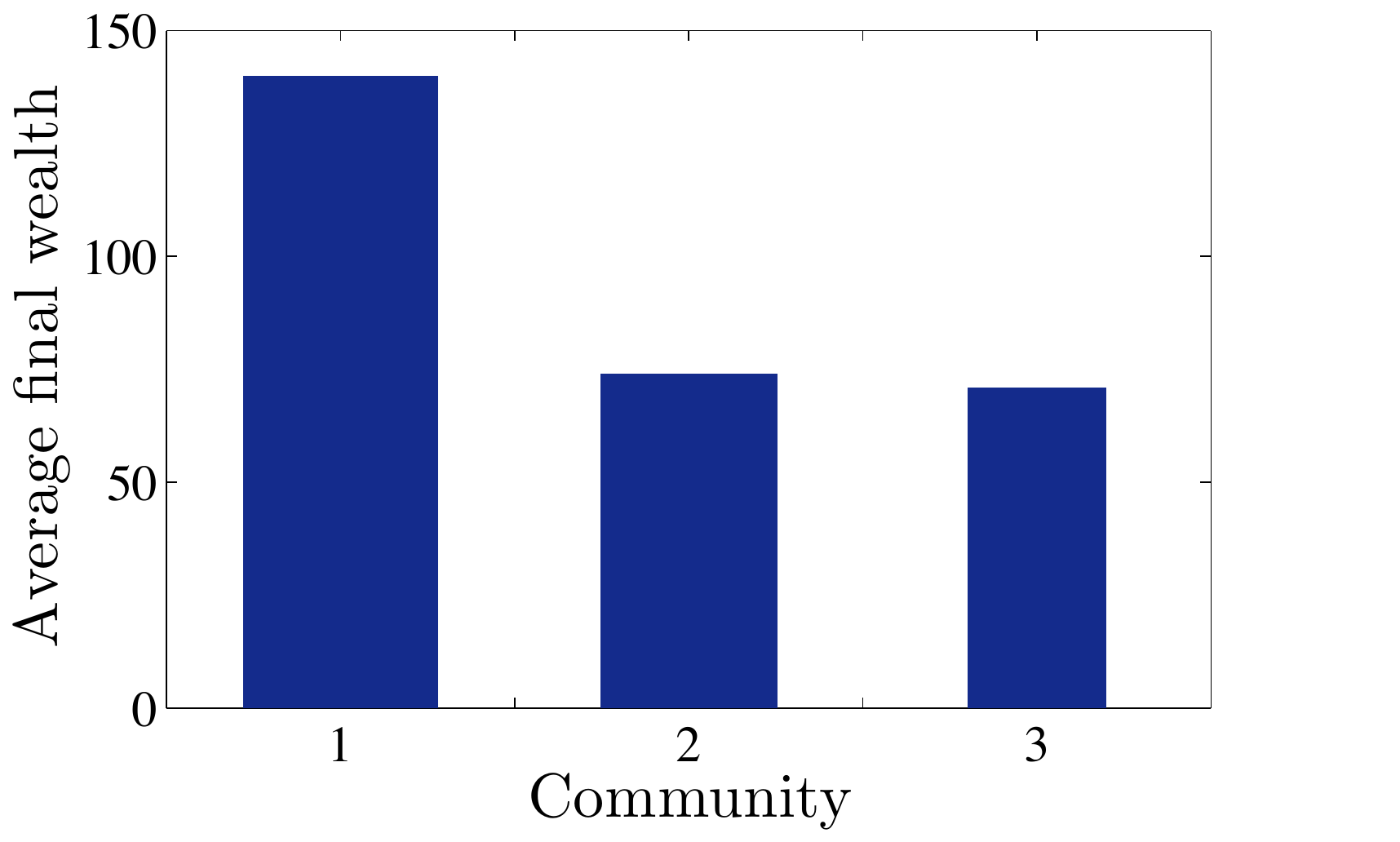}}
\caption{Average final wealth of the agents belonging to communities 1, 2 and 3, when TT scheme is introduced: reference \subref{Fig5a} and focal \subref{Fig5b} scenarios, respectively.}\label{Fig5}
\end{figure}

\subsection{{\bf Focal scenario}} We assume that, after $k_t$ time steps in which the agents invest based on their own risk attitude, the emulation dynamics described in (\ref{eq:5}) are triggered. The triggering instant of the emulation behaviour is indifferent to our purpose, as alternative values of $k_t$ only affect the size of the communities, see figure \ref{Fig9}. The ten richest agents (the leaders) are assumed to have only outgoing edges; the followers, instead, have bidirectional edges with their neighbouring peers, and may have ingoing edges from the leaders. To model this interaction mechanism, we build a directed Barab{\'a}si-Albert (BA) scale-free network \cite{baal:99}, in which the hubs coincide with the leaders. The network is split into three communities, led by the prudent, ordinary, and audacious leaders, respectively. The size of each community is proportional to the sum of the leaders' wealth, which is an indirect measure of their influence.

The results are compared against a twin set of simulations in the reference scenario, sharing the same set of realizations of the Bernoulli variable $B$ in (\ref{eq:2}). For the sake of clarity, we analyse the two taxation schemes separately, starting with the TT scheme. From figure \ref{Fig4}, we observe that this taxation system, regardless of the interactions among the agents, recompenses the prudent strategies in the long run. The emulating the rich interaction, instead, skews the distribution among the communities, see figure \ref{grafi_v2}. In absence of leaders, the interaction with the neighbours would tend to average the agents' attitudes towards risk. However, the presence of leaders differentiates the communities. In particular, the community guided by the prudent leaders preserves a significant number of prudent agents (the 20.29\% of the total population of the community, see table \ref{table1}). Consequently, the average wealth of the agents in the first community is considerably higher compared with the other two communities (see figure \ref{Fig5}\subref{Fig5b}), whose leaders pursue risky strategies which turn to be unprofitable in the long term (their steady-state capital is lower than the average agents' wealth $\bar{x}=100$, see table \ref{table2}). %This effect is also confirmed by looking at the average wealth of the richest agents across the three communities, depicted in.
%
%
%\begin{figure}
%\centering
%\includegraphics[width=7cm]{fig6}
%\caption{Average total wealth of the leaders of the communities 1 (green line), 2 (red line) and 3 (blue line), when a TT is introduced.}\label{Fig6}
%\end{figure}
%
%\begin{figure}
%\subfigure[]{\label{Fig7a}\includegraphics[width=7cm]{fig7a}}
%\subfigure[]{\label{Fig7b}\includegraphics[width=7cm]{fig7b}}
%\caption{Average wealth of the richest agent of the communities 1 (green line), 2 (red line) and 3 (blue line), when a TT is introduced: stubborn agents \subref{Fig7a} and emulating the rich \subref{Fig7b} scenarios, respectively.}\label{Fig7}
%\end{figure}
%
%

\begin{figure}
\centering
\includegraphics[width=6.7cm]{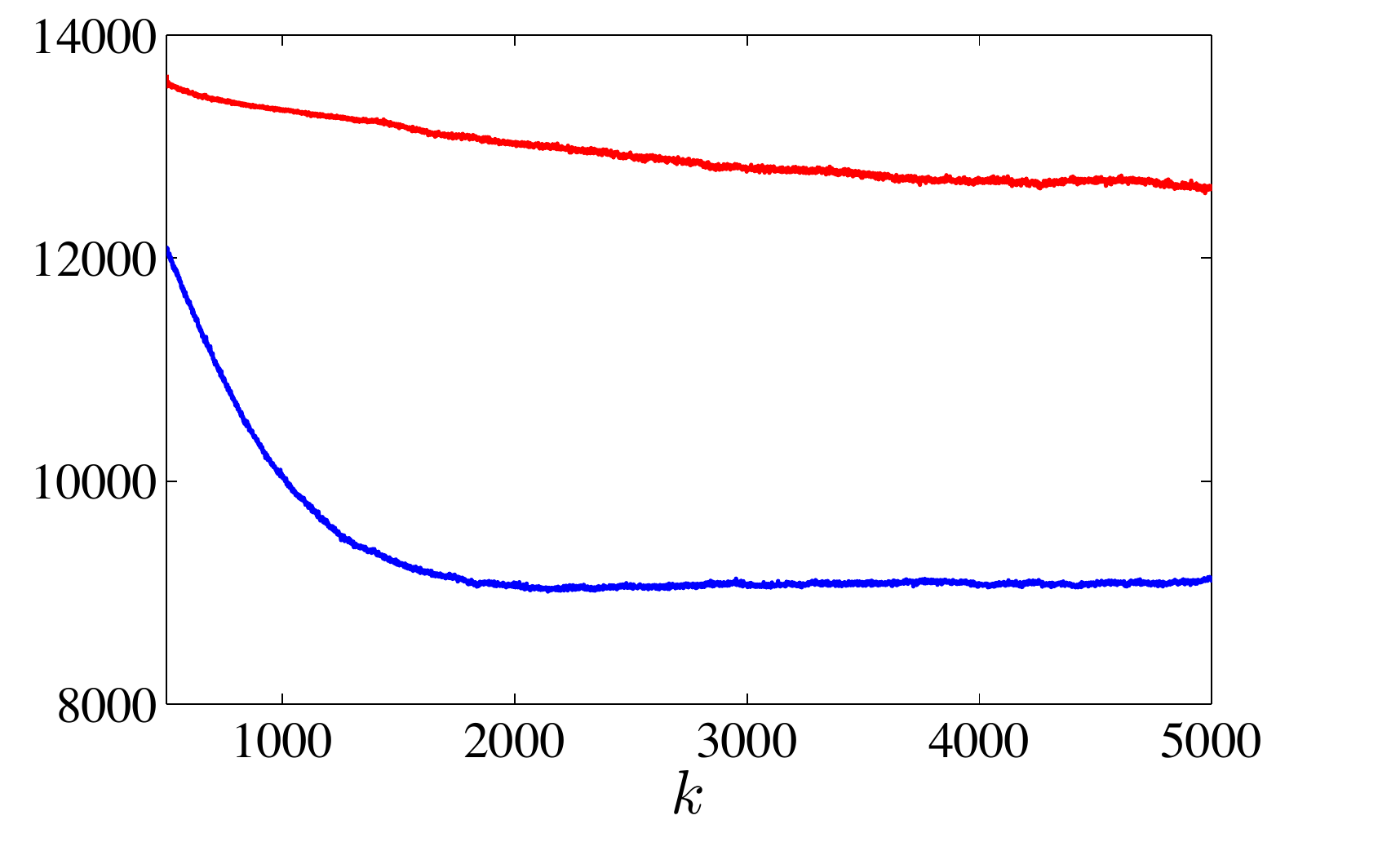}
\caption{Trading volumes when TT scheme is introduced: reference (blue line) and focal (red line) scenarios, respectively.}\label{Fig8}
\end{figure}

Moreover, figure \ref{Fig8} shows that the herding mechanism has the further effect of mitigating the decrease in trading volumes typical of the TT case. This is due to the reduction of the total number of prudent agents illustrated in figure \ref{Fig4}.
%\begin{figure}
%\centering
%\includegraphics[width=7cm]{fig11}
%\caption{Average wealth of the agents belonging to community 2 (red line) and 3 (blue line) when the WT is introduced in the emulating the rich scenario.}\label{Fig11}
%\end{figure}
Differently from the TT case, in which the prudent agents slowly but relentlessly take the leadership of the market, see figure \ref{Fig9}\subref{Fig9a}, when the flat tax is introduced, a prudent strategy is disadvantageous both in the short and in the long term, see figure \ref{Fig9}\subref{Fig9b}. Consequently, no agent is encouraged to emulate the prudent agents, and the market splits into only two communities, guided by the ordinary and audacious leaders, respectively. Accordingly, there are no notable differences in the average wealth of the two communities, see table \ref{table1}.
We emphasize that, while the emulating dynamics can strongly influence the distribution of the wealth across the communities, the overall wealth distribution is only dictated by the taxation scheme. In particular, the variation of the Gini coefficient induced by the emulation dynamics is an order of magnitude lower than that induced by a change of taxation scheme, for any possible value of the interaction weight $w$ and of the average degree of the connection topology. 

\begin{table}
\centering
\caption{Legend. $\bar x_r(T)$ and $\bar x_f(T)$ are the average final wealth in the reference and focal scenarios, respectively; $\nu_{c_i}$ is the average numerosity of the $i$-th community, $i=1,2,3$; $f_j$ is the final percentage of agents belonging to the $j$-th class, $j=1,2,3$. {Confidence intervals with significance level $0.05$ are also reported.}}\label{table1}
\begin{tabular}{|p{0.12\textwidth}|c|c|c|c|c|c|}
\hline 
\multicolumn{1}{|l}{•}  & \multicolumn{3}{|c}{\scriptsize {\textbf{Tobin Tax}}} & \multicolumn{3}{|c|}{{\scriptsize\textbf{Flat Tax}}} \\ 
\hline 
{\scriptsize {\bf Community}} & {\scriptsize \textbf{1}} & {\scriptsize \textbf{2}} & {\scriptsize \textbf{3}} & {\scriptsize \textbf{1}} & {\scriptsize \textbf{2}} & {\scriptsize \textbf{3}} \\ 
\hline
{\scriptsize $\bm{\bar x_r(T)/\bar x}$} & {\scriptsize 1.01} & {\scriptsize 0.97} & {\scriptsize 0.98} & {\scriptsize 0} & {\scriptsize 1.07} & {\scriptsize 0.97} \\[-6pt] 
{} & {\tiny [0.99, 1.02]} & {\tiny [0.96, 0.99]} & {\tiny [0.97, 1.00]} & {} & {\tiny [1.01, 1.13]} & {\tiny [0.94, 1.01]} \\ 
\hline 
{\scriptsize $\bm{\bar x_f(T)/\bar x}$} & {\scriptsize 1.40} & {\scriptsize 0.74} & {\scriptsize 0.71} & {\scriptsize 0} & {\scriptsize 1.05} & {\scriptsize 0.98} \\[-6pt] 
{} & {\tiny [1.38, 1.42]} & {\tiny [0.73, 0.75]} & {\tiny [0.70, 0.73]} & {} & {\tiny [0.99, 1.11]} & {\tiny [0.95, 1.01]} \\ 
\hline 
{\scriptsize $\bm{\nu_{c_i}}$} & {\scriptsize 380.76} & {\scriptsize 347.63} & {\scriptsize 271.61} & {\scriptsize $\ \ 0\ $ } & {\scriptsize 246.24} & {\scriptsize 753.76} \\[-6pt] 
{} & {\tiny [369.89, 391.01]} & {\tiny [336.54, 357.92]} & {\tiny [260.91, 281.71]} & {} & {\tiny [235.76, 256.19]} & {\tiny [742.78, 763.20]} \\ 
\hline
{\scriptsize $\bm{f_1}${\bf (\%)}} & {\scriptsize 20.29} & {\scriptsize 0.60} & {\scriptsize 0.01} & {\scriptsize 0} & {\scriptsize 0.80} & {\scriptsize 0} \\[-6pt]
{} & {\tiny [19.78, 20.80]} & {\tiny [0.53, 0.66]} & {\tiny [0.01, 0.02]} & {} & {\tiny [0.69, 0.92]} & {} \\ 
\hline 
{\scriptsize $\bm{f_2}${\bf (\%)}} & {\scriptsize 79.71} & {\scriptsize 99.40} & {\scriptsize 93.04} & {\scriptsize 0} & {\scriptsize 99.16} & {\scriptsize 93.13} \\[-6pt]
{} & {\tiny [79.20, 80.21]} & {\tiny [99.32, 99.45]} & {\tiny [92.67, 93.41]} & {} & {\tiny [99.04, 99.27]} & {\tiny [92.95, 93.32]} \\
\hline 
{\scriptsize $\bm{f_3}${\bf (\%)}} & {\scriptsize 0} & {\scriptsize 0} & {\scriptsize 6.95} & {\scriptsize 0} & {\scriptsize 0.04} & {\scriptsize 6.87} \\[-6pt] 
{} & {} & {} & {\tiny [6.58, 7.32]} & {} & {\tiny [0.03, 0.06]} & {\tiny [6.68, 7.05]} \\ 
\hline 
\end{tabular}\par
%\bigskip 
\end{table}

{We remark that considering disconnected communities is an idealization of real-world aggregations, where few weak links may still connect the communities. However, 
all the presented results are robust to the addition of links connecting the communities. This is confirmed by a twin set of simulations in which a small fraction (less than 5\%) of the network edges are rewired following a degree-preserving procedure inspired by the work in \cite{lafo:08}. Considering 95\% confidence intervals, we find that the variations of the results are not statistically significant.}
\begin{table}
\centering
\caption{Legend. $\bar x_l(T)$ is the average final wealth of the leaders in the focal scenario; $\nu_{l_i}$ is the average numerosity of the leaders of the $i$-th community, $i=1,2,3$. {Confidence intervals with significance level $0.05$ are also reported.}}\label{table2}
\begin{tabular}{|l|l|c|c|c|}
\hline
& \textbf{{\footnotesize Leaders}} & \textbf{{\footnotesize L1}} & \textbf{{\footnotesize L2}} & \textbf{{\footnotesize L3}} \\ 
\hline
% & $\bm{\bar x_l(k_t)}$ & 1450 & 1788 & 1831 \\
%\cline{2-5}
& {\footnotesize$\bm{\bar x_l(T)/\bar x}$} & {\footnotesize 7.69} & {\footnotesize 0.66} & {\footnotesize 0.64} \\[-6pt]
\textbf{{\footnotesize Tobin Tax}} & {} & {\scriptsize[7.18,8.19]} & {\scriptsize[0.59,0.72]} & {\scriptsize[0.37,0.90]} \\
\cline{2-5}
& {\footnotesize$\bm{\nu_{l_i}}$} & {\footnotesize 4.30} & {\footnotesize 3.20} & {\footnotesize 2.50} \\[-6pt]
& {} & {\scriptsize[4.20,4.40]} & {\scriptsize[3.10,3.29]} & {\scriptsize[2.41,2.59]} \\
 \hline
% & $\bm{\bar x_l(k_t)}$ & 0 & 3598 & 5271 \\
%\cline{2-5}
& {\footnotesize$\bm{\bar x_l(T)/\bar x}$} & {\footnotesize 0} & {\footnotesize 14.29} & {\footnotesize 14.48} \\[-6pt]
\textbf{{\footnotesize Flat Tax}} & {} & {} & {\scriptsize[12.60,15.98]} & {\scriptsize [13.24,15.71]} \\
\cline{2-5}
 & {\footnotesize $\bm{\nu_{l_i}}$} & {\footnotesize 0} & {\footnotesize 3.03} & {\footnotesize 6.97} \\[-6pt]
 & {} & {} & {\scriptsize[2.94,3.12]} & {\scriptsize[6.88, 7.06]} \\
 \hline
\end{tabular}\par
\end{table}
\section{Discussion}\label{sec:concl}
In this paper, we presented a novel agent-based model of a financial market. The agent behaviour has been modelled according to the Von Neumann and Morgenstern utility theory, which led to the definition of three possible trading strategies. Based on the selected strategy, the agents were classified as prudent, ordinary, or audacious. The proposed model allowed for a numerical study of the interplay between herding-like dynamics, the wealth inequalities, and the redistributive effect of two common alternative taxation schemes. 
Specifically, we considered agents capable of adapting their utility function emulating the strategic behaviour of the successful agents. The emerging wealth distribution and trading volumes were compared against the reference scenario of stubborn agents, who maintain their beliefs regardless of the outcome of their trades. The analysis was carried out for two alternative taxation schemes, introducing a Tobin-like tax and a flat tax.
%Specifically, we considered a Tobin-like tax and a flat tax, whose effects were tested in two different scenarios. In the first, agents were stubborn, and maintained their initial risk attitude and trading strategy. In the second, agents were instead adaptive, and prone to update their risk attitude. 
The emerging features of the market have been investigated through a set of extensive simulations in a market populated by 1000 agents.

The numerical analysis of the reference scenario replicated the well known benefits and drawbacks of the two taxation schemes.
Indeed, we observed a trade-off between wealth redistribution and trading volumes: while the Tobin-like tax has the effect of redistributing the wealth among the agents, but reduces the trading volumes, the opposite happens with a flat tax, which encourages to invest, but dramatically increases the disparity among the agents. Moreover, while the TT scheme favoured the prudent agents investing only in the less risky assets, the FT scheme rewarded the audacious agents, that also consider investing in the riskiest assets. 
In the focal scenario, where the adaptive agents consider adjusting their risk attitude and the consequent trading strategy, we observed a significant impact of the agents interactions on the emerging features of the market. Indeed, the richest agents, recognized as the market leaders, formed separate communities. Notably, we observed that the communities benefit from the presence of leaders with successful trading strategies, and are more likely to increase their average wealth. Moreover, this herding-like behaviour mitigated the reduction of the trading volumes typical of Tobin-like taxes, while preserving its redistributive effect. {Ongoing works are devoted to further investigate the driving factors of herding, explicitly accounting for the effect of external inputs, such as the information gathered from the environment.}

\end{document}